\documentclass[pdflatex,sn-mathphys-num]{sn-jnl}

\usepackage[utf8]{inputenc}
\usepackage[T1]{fontenc}
\usepackage{pifont}   
\usepackage{wasysym}  
\usepackage{newunicodechar}

\newunicodechar{✓}{\ding{51}}   
\newunicodechar{✔}{\ding{52}}   
\newunicodechar{✗}{\ding{55}}   
\newunicodechar{✘}{\ding{55}}   
\newunicodechar{○}{\Circle}     
\newunicodechar{●}{\CIRCLE}     

\DeclareUnicodeCharacter{00A0}{\nobreakspace}          
\newunicodechar{–}{--}           
\newunicodechar{—}{---}          
\newunicodechar{×}{\(\times\)}   
\newunicodechar{−}{-}            
\newunicodechar{→}{\(\to\)}      
\newunicodechar{�}{}             

\usepackage{tikz}
\usetikzlibrary{positioning,fit,decorations.pathreplacing,arrows.meta}
\usepackage{pdflscape} 
\usepackage{rotating} 
\usepackage{graphicx}%
\usepackage{multirow}%
\usepackage{amsthm}%

\usepackage{mathrsfs}%
\usepackage[title]{appendix}%
\usepackage{xcolor}%
\usepackage{textcomp}%
\usepackage{manyfoot}%
\usepackage{booktabs}%
\usepackage{algorithm}%
\usepackage{algorithmicx}%
\usepackage{algpseudocode}%
\usepackage{listings}%
\usepackage[font=small,labelfont=bf]{caption}

\makeatletter
\@ifundefined{allowdisplaybreaks}{\def\allowdisplaybreaks{\relax}}{}
\makeatother

\theoremstyle{thmstyleone}%
\theoremstyle{thmstyletwo}%

\theoremstyle{thmstylethree}%

\begin{document}

\title[Article Title]{The AI Roles Continuum: Blurring the Boundary Between Research and Engineering}


\author{\fnm{Deepak Babu} \sur{Piskala}} \email{prdeepak.babu@gmail.com}




\abstract{The rapid scaling of deep neural networks and large language models has collapsed the once-clear divide between “research” and “engineering” in AI organizations. Drawing on public job descriptions, hiring loops, and org narratives from leading labs and tech companies, we propose the \textit{\textbf{AI Roles Continuum}}: a spectrum in which Research Scientists, Research Engineers, Applied Scientists and Machine Learning Engineers share overlapping mandates. Core competencies i.e distributed systems design, large-scale training and optimization, rigorous experimentation, and publication-minded inquiry are now broadly distributed across titles. Treating roles as fluid rather than siloed shortens research-to-production loops, improves iteration velocity, and strengthens organizational learning. We provide a taxonomy of competencies, map them to common titles, and discuss implications for hiring, ladders, and career development in modern AI enterprises.}




\maketitle

\section{Introduction}
In the era of deep neural networks and large language models (LLMs), the traditional separation between ``research'' roles and ``engineering'' roles in AI has increasingly dissolved. Developing state-of-the-art AI systems now demands hybrid skill sets: researchers are expected to write scalable code, and engineers contribute to algorithmic innovations. Top AI organizations have begun explicitly recognizing this convergence. For example, Anthropic notes that “engineers here do lots of research, and researchers do lots of engineering,” observing that the boundary between the two has “dissolved with the advent of large models” \cite{AnthropicCareers2023}. This position paper explores the emerging AI Roles Continuum, where AI Research Scientists, Research Engineers, Applied Scientists, and Machine Learning Engineers increasingly share overlapping responsibilities in creating, scaling, and deploying advanced AI models.

\section{Hybrid Roles in Leading AI Organizations}
Major AI labs and tech companies exemplify how research and engineering roles are structured to overlap:
\begin{itemize}
\item \textbf{OpenAI:} OpenAI hires Research Engineers alongside Research Scientists, expecting them to both build novel AI systems and advance core research. The job description for a Research Engineer emphasizes “solid engineering skills (e.g. designing and improving a massive-scale distributed machine learning system), writing bug-free ML code, and building the science behind the algorithms”\cite{OpenAI-RE-Job}. OpenAI explicitly notes that the most groundbreaking deep learning results are attained at massive scale, requiring engineers comfortable with large distributed systems\cite{OpenAI-RE-Job}. In practice, OpenAI’s research teams are cross-functional, where engineering expertise is considered key to future AI breakthroughs. 

\item \textbf{Anthropic:} Anthropic deliberately avoids a sharp distinction between researchers and engineers. All technical staff share the single title “Member of Technical Staff,” and the company reports that “all of our papers have engineers as authors, and often as first author”\cite{AnthropicCareers2023}, underscoring that research contributions come from both tracks. Anthropic values both research insight and implementation skill in every hire, reflecting their belief that bold new large-model capabilities emerge from tight researcher–engineer collaboration. This “engineer-researcher” fluidity is a conscious part of Anthropic’s team culture\cite{AnthropicCareers2023}. 

\begin{figure}[t]
  \centering
  \begin{minipage}{1.0\linewidth}
    \centering
\begin{tikzpicture}[font=\small]
  \begin{scope}[xscale=12,yscale=1.2]
    \shade[left color=black!10,right color=black!60,rounded corners]
      (0,0) rectangle (1,0.25);
    \foreach \x/\lbl in {0/Research Scientist,0.33/Research Engineer,0.66/Applied Scientist,1/ML Engineer}{
      \draw[black] (\x,0.25) -- ++(0,0.12);
      \node[align=center] at (\x,-0.25) {\lbl};
    }
    \node[anchor=west] at (0.01,0.5) {Novel ML methods};
    \node at (0.33,0.5) {Scaling \& infra};
    \node at (0.66,0.5) {E2E delivery};
    \node[anchor=east] at (0.99,0.5) {Prod integration};
    \draw[decorate,decoration={brace,amplitude=5pt}] (0,-0.45) -- (0.5,-0.45)
      node[midway,below=6pt]{research $\leftrightarrow$ engineering};
    \draw[decorate,decoration={brace,mirror,amplitude=5pt}] (0.5,-0.45) -- (1.0,-0.45)
      node[midway,below=6pt]{engineering $\leftrightarrow$ delivery};
  \end{scope}
  
\end{tikzpicture}
    \captionsetup{justification=centering,name={Fig.}}
    \caption{Roles Continuum. The Research Scientist (RS), Research Engineer (RE), Applied Scientist (AS), and ML Engineer (MLE) exist on a spectrum—from novel method creation to production integration, rather than as disjoint boxes. The gradient and soft overlaps highlight that teams often span multiple mandates simultaneously, depending on product maturity and organizational needs.}
    \label{fig:my-tikz}
  \end{minipage}
\end{figure}
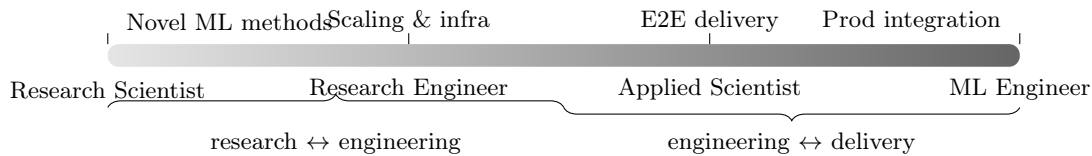

\item \textbf{DeepMind (Google DeepMind):} DeepMind employs Research Engineers who “lead [their] efforts in developing and scaling novel algorithmic methods that push the frontier of AI”\cite{DeepMind-RE-2025}. These roles blend algorithmic innovation with heavy engineering: DeepMind looks for “creative thinkers with strong algorithm-design and programming skills,” including experience in machine learning architectures, distributed training, and optimizing throughput of large models\cite{DeepMind-RE-2025}. A DeepMind Research Engineer is responsible for “design[ing], implement[ing], scal[ing] and evaluat[ing] models and software prototypes,” and for accelerating research by building robust data and training pipelines\cite{DeepMind-RE-2025}. They work closely with Research Scientists; as one DeepMind listing explains, Research Engineers develop “scalable software infrastructure to support the development of novel algorithms and architectures” in pursuit of breakthroughs\cite{DeepMind-RE-2025}. This close partnership blurs the line between who invents an idea and who implements it, since both roles contribute to both code and research. 

\item \textbf{Meta (Facebook AI):} Meta AI tends to use the title Research Scientist even for roles that involve significant engineering. In practice, many “Research Scientists” at Meta engage in software development and infrastructure. As one new hire describes, “Meta hires research scientists to do both research and SWE [software engineering] roles; what they actually do depends on the team”\cite{Lakshman2025-MetaRS}. For example, a Research Scientist (PhD) in Meta’s Systems \& Infrastructure research track noted that the interview process was essentially a standard software engineering interview, and the role was “more similar to a mid-level SWE position,” expecting the scientist to “primarily write software for backend/tooling.”\cite{Lakshman2025-MetaRS} Meta also has Research Engineers in certain teams (e.g. generative language teams), which are strong software engineers with deep AI knowledge\cite{MetaCareers-GenAI-2025}. Overall, Meta’s AI teams are often cross-functional, and an individual’s contribution might span cutting-edge research (e.g. developing a new model technique) to engineering (e.g. coding the training pipeline or deploying the model in a product). The exact balance varies by team, but the company culture encourages researchers to be hands-on with code and engineers to be involved in experimentation. 

\item \textbf{Amazon:} Amazon’s AI-related roles explicitly mix research and engineering duties under titles like Applied Scientist and Research Scientist. According to Amazon’s AI leadership, “research and applied scientists are expected to have deep expertise in a data-driven science discipline … [and] development of software code is a core skill expected from applied scientists as they are deeply involved in bringing their algorithms to production.”\cite{Zeng2022-AmazonCareer} In other words, an Applied Scientist at Amazon is both inventor and engineer, taking models from initial idea through to running at Amazon scale. Job postings for Applied Scientists often require proficiency with big data tools and distributed systems. For instance, one listing highlights experience with “popular deep learning frameworks such as MXNet and TensorFlow” and “large scale distributed systems such as Hadoop, Spark” as preferred qualifications\cite{AIJobs-Applied-Ads-2023}. Key responsibilities of Amazon scientists include designing new ML models and “work[ing] side by side with our engineers to deliver code changes [to] high throughput production systems,” handling very large datasets; even junior Applied Scientist positions mention “working with distributed systems of data” and partnering with product teams to build ML solutions\cite{AmazonAppliedSciJob}. Amazon’s use of the title “Applied Scientist” thus denotes a hybrid job that advances the state of the art in ML while also ensuring the solutions are implemented and scalable on AWS infrastructure. 

\item \textbf{Microsoft:} Microsoft bifurcates between its pure research division (Microsoft Research, where roles are titled Researcher) and product-driven AI roles titled Applied Scientist or ML Engineer in teams like Azure AI or Office. In the product teams, an Applied Scientist is explicitly a cross-disciplinary role. For example, a job posting for Office AI describes “a leadership role at the intersection of applied machine learning, large language models, and multi-modal intelligence” – the scientist is expected to help define product features while leveraging the latest research. Another Microsoft listing for an Applied Scientist II (OneNote team) states the mission is “to deliver high quality AI scenarios,” which involves “adopting the latest ML technologies” and measuring model impact in a real product. These scientists work in tandem with software engineers to integrate models (e.g. Copilot features) into user-facing applications. Meanwhile, Microsoft Research’s Researcher positions focus on foundational research and often require a PhD and publication record, but even in MSR the expectation is that researchers can prototype their ideas (usually in code) and collaborate with engineering teams to transfer successful research into products. Microsoft’s introduction of large-scale AI features (such as GPT-powered Copilots) has led to very tight collaboration between research scientists, applied scientists, and engineers – effectively a continuum of roles all contributing different expertise to the same AI deployment goals. 
\end{itemize}

\noindent Across these organizations, we see a clear trend: roles that were once distinct are converging. Research Scientists are often coding at scale and worrying about deployment, while engineers (Research Engineers, ML Engineers) are contributing to algorithm design and even co-authoring papers. Titles may differ by company, but the hybrid expectations are consistent. As OpenAI aptly put it, “engineering [plays] a key role in most major advances in AI”\cite{OpenAI-RE-Job} – thus everyone on the AI team is expected to bring some engineering prowess \emph{and} some research mindset to the table.

\section{Shared Technical Skills and Responsibilities Across Roles}
Because of this convergence, there is a core set of technical skills and responsibilities now commonly shared across research and engineering roles in AI teams:
\begin{figure}[t]
  \centering
  \begin{minipage}{0.8\linewidth}
    \centering
    \includegraphics[width=\linewidth]{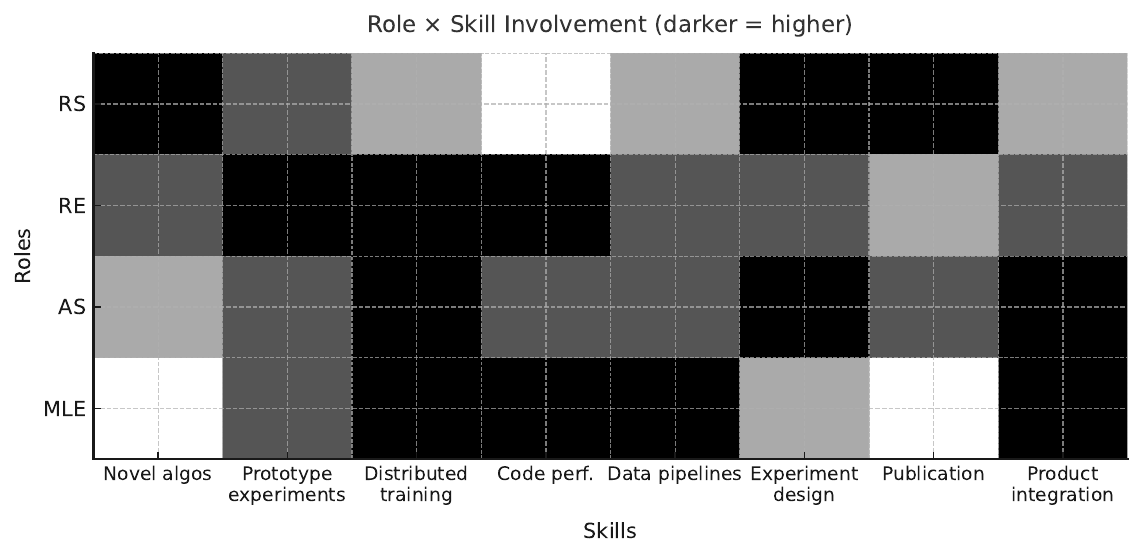}
    \captionsetup{justification=centering,name={Fig.}}
    \caption{Role × Skill involvement heatmap. Darker cells indicate higher typical involvement by a role in that skill (none→strong). The visualization emphasizes overlaps and complementary strengths across Research Scientists (RS), Research Engineers (RE), Applied Scientists (AS), and ML Engineers (MLE), making shared competencies and role-specific focus areas apparent at a glance.}
    \label{fig:role-skill-heatmap}
  \end{minipage}
\end{figure}
\begin{itemize}
\item \textbf{Distributed Systems and Scalable Infrastructure:} The ability to design and work with distributed computing systems is essential for both researchers and engineers building advanced models. Training modern deep networks (like billion-parameter Transformers) requires splitting work across many GPUs/TPUs and machines. For example, OpenAI’s Research Engineer role explicitly seeks experience with “massive-scale distributed machine learning systems”\cite{OpenAI-RE-Job}. Likewise, Amazon Applied Scientists are expected to be comfortable with large-scale distributed systems (e.g. Hadoop, Spark) in addition to ML theory. Research Engineers at DeepMind are hired to “develop scalable software infrastructure” for novel algorithms\cite{DeepMind-RE-2025}, and need familiarity with distributed training and inference infrastructure. In practice, regardless of title, an AI practitioner working on cutting-edge models must understand parallelism \citep{shoeybi2019megatron,rajbhandari2020zero}(data and model parallel training), cluster compute architecture, and possibly cloud platforms or HPC tools. This ensures that ideas can be tested and deployed at the scale where state-of-the-art results emerge. 

\item \textbf{Model Architecture \& Algorithmic Understanding:} Deep knowledge of ML model architectures (e.g. Transformer networks, CNNs, reinforcement learning algorithms) is no longer the sole domain of “researchers.” Engineers implementing these models need to understand how they work to optimize and debug them. Conversely, researchers proposing new architectures often implement them personally. Many job postings reflect this overlap: DeepMind expects its Research Engineers to have expertise in machine learning architectures and algorithm design\cite{DeepMind-RE-2025}, while Amazon’s scientist roles require understanding of “modern methods such as deep learning” to tackle real-world problems. In short, both role types are expected to grasp the theory behind models (to contribute novel ideas or improvements) \emph{and} the practice of coding them. A Research Scientist might come up with a new optimization for Transformers, but a Research/ML Engineer will help integrate it and both need to speak the language of architectures, loss functions, and statistical evaluation. 

\item \textbf{Training Optimization and Efficiency:} Pushing the frontier with large models often hinges on making training faster, more memory-efficient, and more robust. Skills in GPU programming, distributed training optimizations (like gradient accumulation, mixed precision, memory optimization, efficient data pipelines) are in high demand across roles. OpenAI values “high-performance implementations of deep learning algorithms” as a plus even for research-oriented hires\cite{OpenAI-RE-Job}. DeepMind similarly looks for folks who have experience “scaling models [and] optimizing throughput” in their ML systems\cite{DeepMind-RE-2025}. Solving such problems requires both low-level engineering (systems knowledge) and high-level understanding of the learning process. For instance, consider the challenge of speeding up distributed training: one might improve the all-reduce algorithm for gradient synchronization (a systems solution) or adjust the ML algorithm to require less frequent synchronization (an algorithmic solution). As one research engineer noted, “sometimes the right thing to do might be a computer science solution… in a different situation, the solution might be a machine learning one (e.g., update parameters less often and correct for it with importance sampling). The ability to reason about the tradeoffs between these options is what makes a research engineer valuable.”\cite{Matsukawa2019} This ability to balance systems optimization and ML innovation is increasingly expected from both researchers and engineers on large-scale AI projects.

\item \textbf{Machine Learning Frameworks and Tooling:} Proficiency with ML development frameworks (TensorFlow, PyTorch, JAX, etc.) and related tools is universally expected. Research Scientists use these frameworks to prototype new models, while Research/ML Engineers use them to implement, optimize, and deploy models. DeepMind’s job requirements explicitly list expertise in frameworks like JAX and familiarity with the “training/inference infrastructure” around them\cite{DeepMind-RE-2025}. Amazon looks for similar skills (e.g. MXNet, TensorFlow, PyTorch) in applied scientists\cite{AIJobs-Applied-Ads-2023}. In addition, knowing how to use tooling for experiment tracking, evaluation, and deployment (from Jupyter notebooks to Kubernetes, depending on the role) is part of the shared skill set. Everyone on the team might be expected to contribute to, say, a shared PyTorch codebase or to debug a training run on a cluster. Thus, the notion that a researcher only writes research code and an engineer only handles “production code” is fading – both are often using the same tools to write reliable, scalable code (with appropriate testing and optimization) that will eventually become part of a product or research platform. 

\item \textbf{Data Pipeline and Experimentation Skills:} Managing large-scale datasets, data preprocessing, and feeding data efficiently into training is a responsibility often shared across the continuum of roles. An ML Engineer might focus on building a robust data pipeline, but a Research Scientist must also understand data issues to ensure experiments are valid. Many Research Engineers end up creating or improving internal tools for data handling and model evaluation to accelerate research. For example, building evaluation suites or handling logging/monitoring for large experiments can fall to a research engineer, but the scientist using it also needs to know how it works. In summary, both roles engage in the “plumbing” of ML experiments – loading data, choosing metrics, running evaluations – especially with the massive datasets used for deep learning. 

\item \textbf{Cross-Functional Collaboration:} Because deploying an AI model from idea to product involves many stages, cross-functional teamwork is critical. The modern AI team often includes Research Scientists, Applied Scientists, Software Engineers, Data Engineers, and Product Managers working in concert. Therefore, all roles must have collaboration and communication skills and some awareness of the others’ domains. Amazon highlights that its scientists “provide technical partnership to … teams and organizations building machine learning solutions” – meaning an applied scientist might coordinate closely with product engineers or data platform teams. At the same time, engineers on AI teams participate in research discussions, helping to shape research directions with practical insights (e.g. how a model might be integrated into a real system). Many companies actively encourage this fluid collaboration. For instance, Amazon scientists are expected to “partner with … Software Development Engineers” and even mentor them in machine learning. DeepMind’s research engineers “work closely with other engineers and scientists to advance the field”\cite{DeepMind-RE-2025}, and Anthropic’s unified technical staff structure fosters constant interchange between “research” and “engineering” perspectives. The result is that responsibilities like model deployment (traditionally an engineering task) and scientific experimentation (traditionally a research task) are often shared efforts. A Research Scientist might help debug a scaling issue in deployment, while a Research Engineer might co-author a paper about the results – each contributing where needed to achieve the team’s goals. 

\item \textbf{Research Mindset and Publication:} Interestingly, many engineering-focused roles now also value a research mindset and even publication experience. It’s not uncommon for Research Engineers or Applied Scientists to be co-authors on major research papers – in fact, some breakthrough papers (like “Attention Is All You Need” and OpenAI’s GPT-3 technical report) had both engineers and scientists as key contributors. Anthropic explicitly notes that engineers have been first authors on their papers\cite{AnthropicCareers2023}. Companies like Amazon and Microsoft encourage applied scientists to publish in top academic conferences. The rationale is that those in engineering roles are often doing novel work (implementing a new scaling method or improving a model’s efficiency) that is valuable to the research community. Conversely, researchers are encouraged to consider real-world impact, which means writing production-quality code and perhaps releasing datasets or tools. This mutual infusion of research and practical rigor elevates the whole team’s output. In summary, contributing to scientific literature, staying up-to-date with new findings, and thinking in terms of research impact are now part of many engineering roles – just as research roles are expected to produce robust, working systems, not just papers. 
\end{itemize}

\section{Role Definitions and Skill Requirements in Practice}
To illustrate the above points, we can examine the expectations for several common AI job titles. Real-world job postings and descriptions from leading AI employers highlight both the distinct focuses and the overlapping skills of each role:

\begin{itemize}
\item \textbf{Research Scientist (RS):} This title often denotes someone focused on pioneering new algorithms or fundamental research questions. Typical requirements include an advanced degree (usually a Ph.D.) in a relevant field and a strong publication record in top-tier conferences. Research Scientists are expected to have deep expertise in areas like machine learning theory, NLP, computer vision, etc., and to generate original research ideas. However, they are also usually competent programmers; they must implement experiments (generally in Python with ML libraries) to validate their ideas. In industry labs, an RS is often also responsible for moving their models toward production. For example, at Meta, a Research Scientist might spend a significant portion of time coding and even undergo coding interviews to get hired\cite{Lakshman2025-MetaRS}. At Amazon, Research Scientists fall under the broader science career track and are evaluated on both scientific inventiveness and the ability to “apply scientific principles to support significant invention” in real products\cite{Zeng2022-AmazonCareer}. In summary, while an RS emphasizes research novelty, their skill set often spans from mathematical derivations to writing efficient code. They might lead projects, author papers, and prototype new models and in many cases will hand off robust implementations to engineers for scaling, unless they choose to scale it themselves. 

\item \textbf{Research Engineer (RE):} A Research Engineer is an archetypal hybrid role. Found at organizations like OpenAI, DeepMind, Anthropic, etc., REs are strong software engineers who also have a solid understanding of ML. They typically do not require a Ph.D. (industry experience or a Master’s can suffice), but they are expected to be excellent in coding, systems design, and running experiments. An RE’s primary responsibility is often described as “enabling, contributing to, and accelerating ML research by bringing engineering expertise to the projects.” In practice, this means REs implement new algorithms (turning paper ideas into running code), set up large-scale training runs, manage model codebases, and build tooling for researchers (such as evaluation pipelines or visualization dashboards). They are the ones who, for instance, ensure that a new model architecture can actually be trained across dozens of GPUs without crashing, or that an idea gets tested on proper datasets. But beyond support, Research Engineers frequently contribute new ideas and improvements based on their empirical findings. Teams often have REs and Research Scientists working as peers – it’s noted that “research scientists and research engineers… both come up with ideas and implement those ideas,” and they can be “equal contributors to papers.”\cite{Huyen2022} The main difference is bureaucratic (scientists might have higher rank or pay) or experiential (scientists might have more theory background), but day-to-day, an RE might be writing code one day and brainstorming algorithm improvements the next. Key skills for REs include strong programming (C++/Python), knowledge of ML frameworks, ability to set up distributed training jobs, debugging numerical issues, and a good intuition for both software performance and model behavior. This role exemplifies the blurred line: it sits squarely between pure engineering and pure research, and different companies place it slightly closer to one side or the other. (For instance, DeepMind’s REs might lean slightly more toward engineering for novel research, while OpenAI’s REs might directly train giant models and thereby engage in research discovery as well.) 

\item \textbf{Applied Scientist / Applied Researcher:} Titles like Applied Scientist (used at Amazon, Microsoft, and others) indicate a role focused on applying and translating AI research into real-world products or services. These roles typically require strong expertise in ML (often a Ph.D. or equivalent experience) and the ability to write production-quality code. As Amazon describes, an Applied Scientist is someone who “bridges the gap between research and engineering by taking state-of-the-art models and making them work at scale for customers.” They work on product-oriented teams (e.g. Alexa, AWS AI services, personalization, ads) and are responsible for the end-to-end development of ML solutions: from formulating the model, training it on large datasets, to deploying it in the production environment. An Applied Scientist might one day be reading research papers or inventing a new modeling approach, and the next day be optimizing latency in an AWS service or A/B testing a feature. Job postings frequently list a mix of skills: deep learning, distributed data processing, programming in a language like Python or Java, and the ability to evaluate business impact. At Amazon, they are also expected to adhere to Amazon’s leadership principles, showing business insight and customer focus in their research decisions. Microsoft’s Applied Scientists similarly work on teams like Bing, Office, or Azure, where they implement models (say, an LLM for Office Copilot) and measure how it improves user experience. In summary, the Applied Scientist is essentially both a researcher (coming up with novel solutions) and an engineer (bringing those solutions to production fruition), with perhaps more emphasis on delivering customer-facing impact than publishing papers. They often collaborate with software engineers who integrate their models into larger systems, but the scientists themselves ensure the models are correctly built and tuned for production use. 

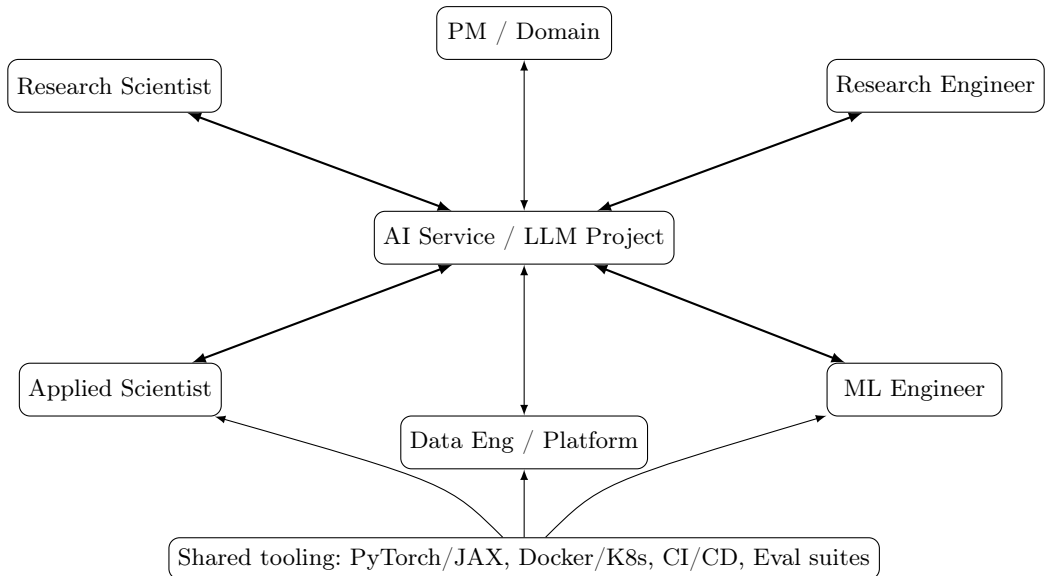
\begin{figure}[t]
  \centering
  \begin{minipage}{1.0\linewidth}
    \centering
\begin{tikzpicture}[>=latex,font=\small]
  \tikzstyle{role}=[rectangle,rounded corners,draw,minimum width=23mm,minimum height=7mm,align=center]
  \node[role] (core) {AI Service / LLM Project};

  \node[role,above left=1.3cm and 2.0cm of core] (rs) {Research Scientist};
  \node[role,above right=1.3cm and 2.0cm of core] (re) {Research Engineer};
  \node[role,below left=1.3cm and 2.0cm of core] (as) {Applied Scientist};
  \node[role,below right=1.3cm and 2.0cm of core] (mle){ML Engineer};

  \foreach \n in {rs,re,as,mle}{
    \draw[<->,thick] (\n) -- (core);
  }

  \node[role,below=2.0cm of core] (data) {Data Eng / Platform};
  \node[role,above=2.0cm of core] (pm) {PM / Domain};
  \draw[<->] (data) -- (core);
  \draw[<->] (pm) -- (core);

  \node[draw,rounded corners,align=center,below=0.9cm of data,minimum width=70mm]
       (tools) {Shared tooling: PyTorch/JAX, Docker/K8s, CI/CD, Eval suites};
  \draw[->] (tools) -- (data);
  \draw[->] (tools) .. controls +(1,1) .. (mle);
  \draw[->] (tools) .. controls +(-1,1) .. (as);
\end{tikzpicture}
    \captionsetup{justification=centering,name={Fig.}}
    \caption{Cross-functional team topology. An AI/LLM project sits at the core, with RS, RE, AS, and MLE connected bidirectionally, and adjacent partners (PM/Domain, Data Engineering/Platform) interfacing through shared tooling (e.g., CI/CD, K8s, training/eval stacks). The diagram emphasizes clear ownership with porous boundaries that enable rapid iteration.}
    \label{fig:team-topology}
  \end{minipage}
\end{figure}

\item \textbf{Machine Learning Engineer (ML Engineer):} This title often refers to a role more rooted in the engineering side, especially in organizations outside of core research labs. An ML Engineer is typically a software engineer who specializes in implementing and deploying machine learning models. Key skills include building data pipelines, serving models in production (e.g. via APIs or embedding into applications), and optimizing runtime performance (CPU/GPU inference optimization, memory management, etc.). They might also handle continuous training pipelines, monitoring of model performance in production, and A/B testing frameworks. While an ML Engineer may not be expected to invent new ML algorithms from scratch, they do need to understand ML well enough to choose appropriate models and troubleshoot them. In many companies, ML Engineers work closely with Data Scientists or Research Scientists: the scientist might develop a model in a notebook, and the ML Engineer will refactor that into robust code, integrate it with production data sources, and ensure it scales and meets latency requirements. However, as the continuum blurs, many ML Engineers today also experiment with model tuning and contribute to research-oriented tasks (like feature engineering or even writing custom loss functions), and conversely, some research scientists in product teams end up deploying models themselves. For example, at Meta, the term “ML Engineer” can sometimes be used interchangeably with Research Engineer or even Software Engineer on AI teams, indicating a role that requires strong coding skills \emph{plus} ML knowledge. Generally, ML Engineers excel at translating a proven idea into a reliable service. Their toolkit might include TensorFlow Serving or TorchServe, pipeline tools like Apache Beam or AWS SageMaker, and good software development practices. While the pure ML Engineer role might not be writing academic papers, it is vital for bridging the last mile from an AI model to a usable product and often requires creative problem-solving that overlaps with research (for instance, figuring out how to compress a large model for mobile deployment can involve researching model quantization\citep{dettmers2022llmint8,frantar2022gptq,dettmers2023qlora} or distillation\citep{hinton2015distill} techniques). 
\end{itemize}

\noindent All these roles share a considerable intersection in skills. For instance, both Research Engineers and ML Engineers need to know how to handle distributed training; both Research Scientists and Applied Scientists need to stay current with literature and also code in frameworks like PyTorch. Many organizations acknowledge that an employee’s title doesn’t rigidly define what they do day-to-day. It’s common to see a Research Scientist at a company like Google or Meta spending a big chunk of time on coding and infrastructure, while a Research Engineer or ML Engineer might be driving a research initiative. As one analysis put it, “Research scientists should, first and foremost, be engineers” in practice\cite{Huyen2022} – meaning that even those hired for research are expected to implement and experiment, not just theorize. And conversely, engineering-focused staff are often contributing ideas and improvements that shape the research direction. 

\section{Scaling AI Infrastructure and Cross-Functional Team Structures}
One of the driving forces behind the roles continuum is the sheer complexity of modern AI projects. Training and deploying frontier models (such as GPT-4, PaLM, or Claude) is a socio-technical endeavor that requires diverse expertise. The infrastructure for these models – distributed training clusters, model parallelism strategies, massive datasets, continuous integration of model updates – cannot be managed by traditional siloed teams. Industry leaders have adopted cross-functional team structures for AI, which further blur role boundaries.

In cross-functional AI teams, individuals with different titles collaborate closely throughout the project lifecycle. A concrete example is how AI model scaling is tackled. Consider a team working on a new large language model: researchers may propose model improvements, but research engineers will write the code to try them, and ML engineers will profile and optimize the training process. They must work in tandem to iterate quickly. As Akihiro Matsukawa \cite{Matsukawa2019} noted in his \textit{Research Engineering FAQs}, scaling a large-model training job might present a choice between a systems solution (better network algorithms, memory management) and a modeling solution (algorithmic tweaks to use resources more efficiently), and deciding the best approach “requires reasoning about tradeoffs” across domains\cite{Matsukawa2019}. This kind of decision can only be made when team members understand each other’s domains – e.g., the researcher understands some systems constraints and the engineer understands some ML theory. Thus, teams often sit together, use the same tools, and have fluid communication loops.

Another aspect is model deployment and product integration. Deploying an AI model (especially an LLM powering a product like ChatGPT or an AI Copilot) is not a hand-off where research throws a model over the wall to engineering. Instead, deployment is iterative and collaborative. Research engineers and ML engineers set up serving infrastructure, but also need feedback from research scientists on how to handle model updates or evaluate outputs for quality. Likewise, scientists may need to be aware of constraints discovered in deployment (like memory limits or real-time user input patterns) which could inspire research into more efficient architectures or prompt-tuning methods. Companies like Amazon and Microsoft often embed scientists directly into product teams so that such feedback loops happen daily. The result is that roles intermingle: a software engineer might suggest a change in the model to improve latency, or a researcher might suggest an engineering change (such as a different distributed computing strategy) to allow exploring a bigger model. The team as a whole owns the outcome (a working, high-performing AI service), and each member extends their skills as needed to achieve it.

Organizationally, some companies have explicitly broken down the walls between research and engineering departments. OpenAI, for instance, calls itself an “AI research and deployment company” and has many team members with dual skill sets. Anthropic’s single-track technical staff is another explicit example, where hiring and interviews are tailored to find people who can contribute in multiple ways regardless of their background. Google (after merging Brain and DeepMind into Google DeepMind) also emphasizes research-to-production pipelines, often moving people between pure research and applied teams. This fluidity is part of an emerging AI engineering culture that values versatility. New job titles like “Machine Learning Scientist” or “AI Generalist” even pop up in smaller AI startups, indicating an expectation that one person may wear both hats.

From a project management perspective, having blended roles can speed up progress. Instead of a linear process (research $\rightarrow$ handoff $\rightarrow$ engineering $\rightarrow$ handoff $\rightarrow$ deployment), cross-functional teams iterate in parallel. For example, during the development of a large model, as soon as a research insight is found, it can be tested in a realistic setting by the engineers, and if issues arise, researchers can address them quickly. This was essential in landmark projects like OpenAI’s GPT-3 and GPT-4, where novel techniques (e.g. scaling laws \citep{kaplan2020scaling,hoffmann2022chinchilla} or reinforcement learning from human feedback \citep{christiano2017prefs,ouyang2022instructgpt}) had to be tested on enormous infrastructure – requiring tight partnership between those inventing the methods and those running the clusters. Indeed, some of the acknowledgments and reports from these projects mention the critical role of engineering talent in achieving the results (even if not by name due to non-public contributions). As a community, it’s recognized that major advances in AI now often come from engineering breakthroughs (better parallelism, new hardware utilization, etc.) as much as from pure algorithmic breakthroughs.

Hence, research and engineering are viewed as two facets of the same overall innovation process, rather than sequential or separate processes.

\textbf{Team structure trends:} Many organizations maintain a mix of people with PhDs and people with software engineering backgrounds on each team, rather than separating them. They might still have different performance metrics (e.g. researchers might be evaluated on published papers or novel patents, engineers on system reliability and code quality), but the daily work is intertwined. Some companies rotate individuals through different types of projects to broaden their skill sets. Amazon, for example, has programs to help scientists gain engineering skills and vice versa. Microsoft has a concept of “software engineer in ML” vs. “researcher,” but in the new AI-powered products they effectively function together. Startups often ignore titles entirely in favor of “AI Engineer” or similar, expecting end-to-end ownership of both research and engineering tasks.

In terms of infrastructure, the rise of common platforms and tools has also helped blur roles. Using a unified platform (like TensorFlow or PyTorch) for both experimentation and deployment means everyone on the team speaks the same “language” to some extent. Tools like Docker, Kubernetes, or cloud ML services are used by scientists to run experiments and by engineers to deploy – this common toolkit makes it easier for roles to swap tasks or assist each other. When a researcher can containerize their training code, an engineer can directly use that in a production pipeline with minimal translation. This DevOps-style approach to ML (sometimes called MLOps) encourages researchers to think more like engineers in terms of writing maintainable code, and encourages engineers to accommodate the rapid, iterative nature of research.

Finally, cross-functional AI teams also include roles like data engineers, product managers, etc., which further necessitate clear communication. A research-minded person must explain model behaviors or needs to a product manager, and an engineering-minded person must understand the scientific objectives driving a feature. This all contributes to an environment where strict role definitions break down in favor of a continuum of expertise – each team member has a T-shaped skill profile (deep in some area, broad across others) and the overlaps between them ensure there are no “gaps” in turning an idea into a deployed, reliable AI system.

\section{Organizational Demand Along the Research--Engineering Continuum}\label{sec:org-mix}

\begin{figure}[t]
  \centering
  \includegraphics[width=\linewidth]{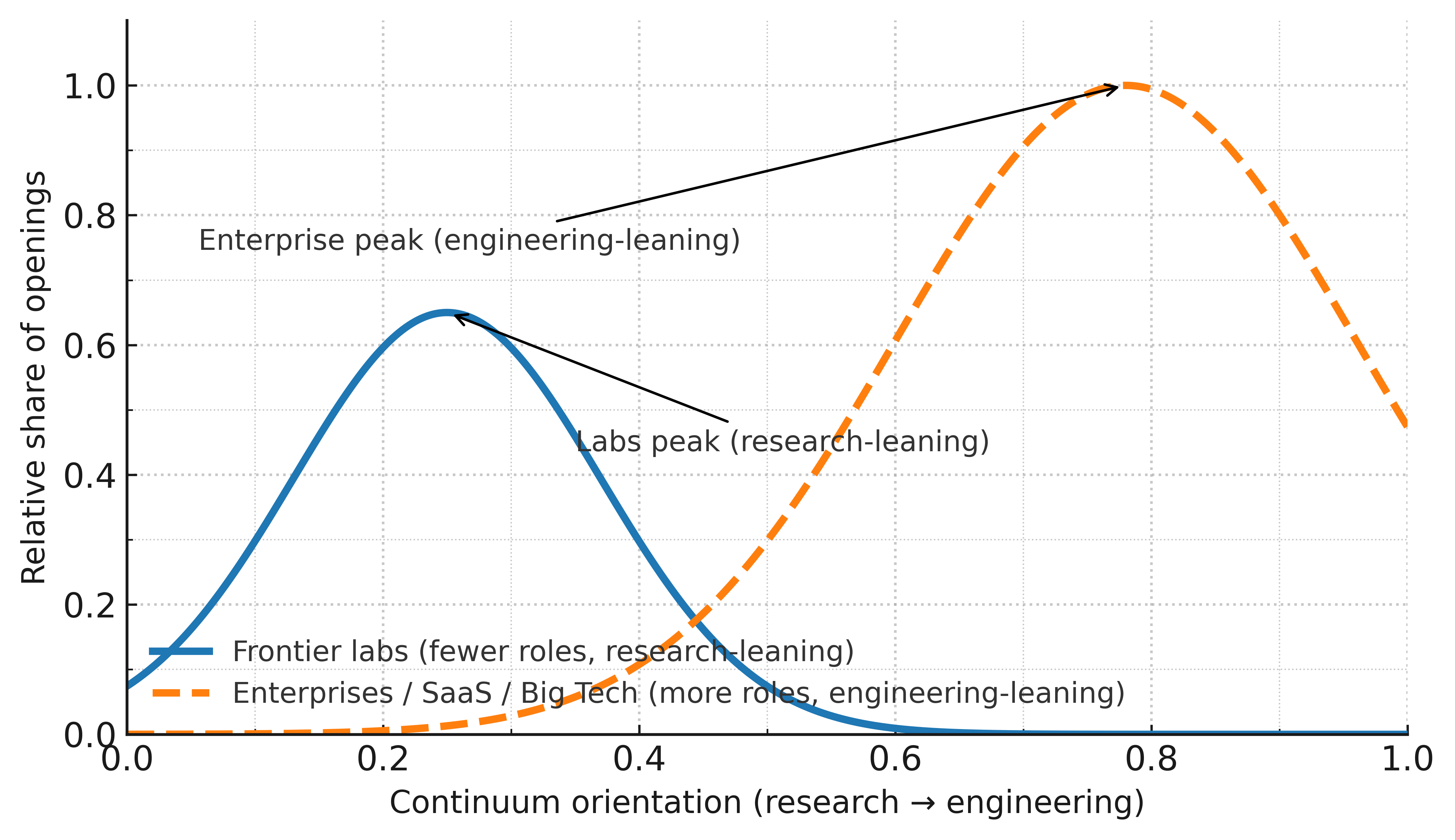}
  \caption{Illustrative hiring distribution along the research--engineering continuum by organization type. Frontier labs peak on the research-leaning side with fewer total openings; enterprises (vertical AI / SaaS / big tech) peak on the engineering-leaning side with more openings.}
  \label{fig:org-continuum}
\end{figure}

The continuum does not make organizations look alike; it \emph{tilts} them. What varies by archetype is where hiring mass settles along the research–engineering axis and why. As Figure~\ref{fig:org-continuum} suggests, frontier labs concentrate a small number of openings on the research-leaning side, while product and platform companies create far more roles that cluster toward engineering. Yet in every setting, the most effective teams are hybrid and mobile along the slider.

Frontier model labs, those building general-purpose foundation models and the science around them, optimize for discovery at scale. Their most frequent hires are research scientists who can originate methods and evaluation schemes, paired with research engineers capable of turning ideas into massive training runs and rigorous ablations. Engineering is omnipresent, but it exists in service of scientific acceleration: cluster orchestration, data curation at scale, and the internal agent stacks used to red-team models, generate synthetic curricula, and probe capability boundaries. Memory abstractions and tool-use are built primarily as \emph{research instruments} here, not as external products.

Foundation-model platforms (often cloud/API providers) sit closer to the center of the continuum. Their mandate is to operate models as reliable services and to expose model-adjacent \emph{primitives} memory, retrieval, tool-calling, safety filters, telemetry as stable APIs. This pushes hiring toward applied scientists who can harden fine-tuning and routing strategies, and toward ML/Platform engineers who can make serving predictable in cost, latency, and quality. Research persists, but it is oriented to efficiency (distillation\citep{hinton2015distill}, quantization \citep{dettmers2022llmint8,frantar2022gptq,dettmers2023qlora}), controllability, and safety. The gestalt is productized science: the same agent \citep{yao2022react,zhou2023webarena,liu2023agentbench,wang2023voyager} and memory ideas used in labs are recast as features that developers can compose.

A distinct archetype has emerged around agent frameworks and developer tooling. These companies treat \emph{agents as a software stack}: planners \citep{yao2022react} and tool \citep{schick2023toolformer} registries, episodic and long-horizon memory \citep{wu2022memorizing,khandelwal2019knnlm}, simulators for evaluation \citep{liang2022helm,hendrycks2021mmlu,srivastava2022bigbench,lin2022truthfulqa}, sandboxes for safe execution, and observability to explain what an agent tried and why. Hiring here still leans engineering, but with a strong seam of applied research in routing, self-reflection, and evaluation-by-design. The work is less about inventing new base models and more about making agentic behavior legible, reliable, and programmable across diverse tools and data backends.

Vertical-AI and SaaS product companies push furthest toward the engineering end of the slider while keeping a pragmatic research thread. They turn general models into domain outcomes i.e customer support resolution, sales assist, IT automation, clinical or financial drafting by layering \emph{memory} (user profiles, case history, policy context) and \emph{tool use} (ticketing, CRM, payments, knowledge systems) around an FM core. The day-to-day work blends prompt/RAG \citep{lewis2020rag} design, guardrail and reward/policy tuning, A/B experimentation, and relentless attention to latency, reliability, and cost. Applied scientists and product-oriented ML engineers dominate headcount; research shows up where the domain demands specialization (speech, document intelligence, regulated text).

Big-tech product organizations resemble vertical SaaS at extreme scale. Their center of gravity is engineering and experimentation: privacy-aware telemetry, cross-application memory, abuse/fraud defenses, policy compliance, and the plumbing that allows agents to orchestrate across operating systems and productivity suites. They still hire researchers, but often for metrics methodology, personalization objectives, and safety science that can survive at population scale. The practical signature is that agents are not a demo, they are integrated into surfaces that millions of people touch.

Enterprise adopters in the Global 2000 emphasize integration, governance, and reliability over invention. Their hiring tilts toward platform engineers for LLMOps, solutions architects who can wire agents into legacy systems, and data/ML engineers who can make retrieval governance and PII-aware memory work under compliance constraints. The research dial turns up only in regulated analytics or where proprietary data afford a method advantage. Services and systems integrators orbit this landscape, translating the same stack into client contexts and scaling change management; their demand profile mirrors enterprise adopters but with stronger delivery cadence and reference architectures for memory, tool-calling, and evaluation.

Finally, open-source ecosystems and academic labs outside the frontier cohort provide the commons: methods, datasets, and reference stacks that make agents, memory, and evaluation reproducible. Their hiring mixes research and engineering in roughly equal measure, enough depth to propose new ideas and enough craft to ship maintainable libraries that others can trust. In practice, these groups set de facto standards for memory schemas, tool interfaces, and evaluation protocols, which then flow back into platforms and products.

In synthesis, \emph{labs} are far fewer by headcount and skew toward science; \emph{platforms, tools, and product companies} are far larger and skew toward engineering. But across all archetypes the same ingredients repeat namely agents, memory, tool use, and evaluation and the premium accrues to people who can slide along the continuum: invent when needed, engineer when it matters, and reason about safety, reliability, and product impact throughout.

\section{Conclusion}

\begin{sidewaystable}[htbp]
\centering
\renewcommand{\arraystretch}{1.2} 
\begin{tabular}{p{4.5cm}cccc}
\hline
\textbf{Skill/Responsibility} & \textbf{Research Scientist} & \textbf{Research Engineer} & \textbf{Applied Scientist} & \textbf{ML Engineer} \\
\hline
Propose novel ML algorithms            & ✔ (primary)       & ✔ (often)        & ✔ (applied focus)   & ○ (occasionally)    \\
Implement models \& experiments        & ✔ (prototyping)   & ✔✔ (primary)     & ✔ (to validate)     & ✔✔ (primary)        \\
Scalable distributed training          & ✔ (know basics)   & ✔✔ (primary)     & ✔✔ (primary)        & ✔✔ (primary)        \\
Code optimization (performance)        & ○ (basic needed)  & ✔✔ (extensive)   & ✔ (practical needed)& ✔✔ (extensive)      \\
Data pipeline \& processing            & ○/✔ (often delegated) & ✔ (some)    & ✔ (some)           & ✔✔ (primary)        \\
Experiment design \& analysis          & ✔✔ (primary)      & ✔ (with team)    & ✔✔ (primary, business metrics) & ✔ (in validation)    \\
Software engineering best practices    & ○/✔ (varies)      & ✔✔ (primary)     & ✔✔ (primary)        & ✔✔ (primary)        \\
Publication \& research community      & ✔✔ (primary output) & ✔ (sometimes co-author) & ✔ (encouraged to publish) & ○ (rare)       \\
Product integration \& deployment      & ○ (support role)  & ✔ (support role) & ✔✔ (primary goal)   & ✔✔ (primary goal)   \\
\hline
\end{tabular}
\caption{Key: ✔✔ = core focus of role; ✔ = frequently involved; ○ = minimally or indirectly involved. Actual responsibilities vary by organization and team.}
\label{tab:roles_matrix}
\end{sidewaystable}
 
The AI Roles Continuum is a conceptual framework for understanding modern AI work: instead of siloed “research vs engineering,” we have a spectrum where most practitioners are hybrid to varying degrees. This shift is evident from the way top AI labs hire (valuing mixed skill sets), how job descriptions are written (emphasizing both algorithm and implementation expertise), and how teams operate (cross-collaboration from idea to deployment). Blurring the boundary between research and engineering has proven to be a competitive advantage in pushing AI’s frontiers – it leads to faster iteration, more robust innovations, and individuals who can navigate the full landscape of AI development. As AI systems scale in complexity and impact, the trend is likely to continue: the best AI professionals will be those who can seamlessly move along the research–engineering continuum, embodying the strengths of both.

The above matrix is a generalized view – in practice individuals often exceed these expectations (e.g., many Research Engineers also publish extensively, and some Research Scientists commit production code). The overarching point is that there is significant convergence, and success in modern AI projects comes from recognizing and leveraging this continuum of roles rather than treating research and engineering as separate silos. 

\textit{Limitations.} Our evidence base relies on public artifacts that can be aspirational and uneven across regions and firm sizes. The continuum is a conceptual synthesis rather than a causal model, and we did not measure outcome differences across organizations. Because titles and responsibilities are evolving rapidly, our claims should be read as time-bounded to 2024–2025. Expanding to longitudinal and cross-regional data, and to anonymized internal ladders and evaluations, is a priority for future work.

\bibliography{sn-bibliography_cleaned}

@misc{AnthropicCareers2023,
  author       = {{Anthropic}},
  title        = {Careers at Anthropic},
  howpublished = {\url{https://www.anthropic.com/careers}},
  note         = {Engineers do research and researchers do engineering},
  year         = {2023},
  urldate      = {2025-07-23}
}

@misc{OpenAI-RE-Job,
  author       = {{OpenAI}},
  title        = {Research Engineer --- Job Description},
  howpublished = {\url{https://openai.com/careers}},
  year         = {2024},
  urldate      = {2025-07-23}
}

@misc{DeepMind-RE-2025,
  author       = {{Google DeepMind}},
  title        = {Research Engineer (AI for Sustainability) Job Posting},
  howpublished = {Built In SF Jobs},
  year         = {2025},
  urldate      = {2025-07-23}
}

@article{Zeng2022-AmazonCareer,
  author  = {Belinda Zeng},
  title   = {How to build a successful career as a scientist at Amazon},
  journal = {Amazon Science},
  year    = {2022},
  month   = {January 26}
}

@misc{AmazonAppliedSciJob,
  author       = {{Amazon.com}},
  title        = {Applied Scientist, Advertising --- Job Listing},
  howpublished = {AIJobs.net},
  year         = {2023},
  urldate      = {2025-07-23}
}

@misc{Lakshman2025-MetaRS,
  author       = {Aidan Lakshman},
  title        = {What is it like to interview at Meta/Facebook? (Research Scientist)},
  howpublished = {Personal blog},
  year         = {2025},
  month        = {February 17},
  urldate      = {2025-07-23}
}

@misc{Matsukawa2019,
  author       = {Akihiro Matsukawa},
  title        = {Research Engineering FAQs},
  howpublished = {\url{https://mtskw.com/posts/re/}},
  year         = {2019},
  month        = {July 7},
  urldate      = {2025-07-23}
}

@misc{Huyen2022,
  author       = {Chip Huyen},
  title        = {Research scientist vs. research engineer},
  howpublished = {\url{https://huyenchip.com/ml-interviews-book/contents/1.1.2.2-research-scientist-vs.-research-engineer.html}},
  year         = {2022},
  urldate      = {2025-07-23}
}

@article{kaplan2020scaling,
  title={Scaling Laws for Neural Language Models},
  author={Kaplan, Jared and McCandlish, Sam and Henighan, Tom and Brown, Tom B. and Chess, Benjamin and others},
  journal={arXiv preprint arXiv:2001.08361},
  year={2020}
}

@inproceedings{hoffmann2022chinchilla,
  title={Training Compute-Optimal Large Language Models},
  author={Hoffmann, Jordan and Borgeaud, Sebastian and Mensch, Arthur and Buchatskaya, Elena and Cai, Trevor and others},
  booktitle={NeurIPS},
  year={2022},
  note={arXiv:2203.15556}
}

@inproceedings{christiano2017prefs,
  title={Deep Reinforcement Learning from Human Preferences},
  author={Christiano, Paul and Leike, Jan and Brown, Tom B. and Martic, Miljan and Legg, Shane and Amodei, Dario},
  booktitle={NeurIPS},
  year={2017},
  note={arXiv:1706.03741}
}

@inproceedings{ouyang2022instructgpt,
  title={Training Language Models to Follow Instructions with Human Feedback},
  author={Ouyang, Long and Wu, Jeff and Jiang, Xu and Almeida, Diogo and Wainwright, Carroll and others},
  booktitle={NeurIPS},
  year={2022},
  note={arXiv:2203.02155}
}

@inproceedings{lewis2020rag,
  title={Retrieval-Augmented Generation for Knowledge-Intensive NLP Tasks},
  author={Lewis, Patrick and Perez, Ethan and Piktus, Aleksandra and Petroni, Fabio and Karpukhin, Vladimir and others},
  booktitle={NeurIPS},
  year={2020},
  note={arXiv:2005.11401}
}

@inproceedings{yao2022react,
  title={ReAct: Synergizing Reasoning and Acting in Language Models},
  author={Yao, Shunyu and Zhao, Jeffrey and Yu, Dian and Du, Nan and Shafran, Izhak and Narasimhan, Karthik and Cao, Yuan},
  booktitle={ICLR},
  year={2023},
  note={arXiv:2210.03629}
}

@inproceedings{schick2023toolformer,
  title={Toolformer: Language Models Can Teach Themselves to Use Tools},
  author={Schick, Timo and Dwivedi-Yu, Jane and Dess{\`i}, Roberto and Raileanu, Roberta and Lomeli, Maria and Zettlemoyer, Luke and Cancedda, Nicola and Scialom, Thomas},
  booktitle={NeurIPS},
  year={2023},
  note={arXiv:2302.04761, DOI:10.5555/3666122.3669119}
}

@inproceedings{wu2022memorizing,
  title={Memorizing Transformers},
  author={Wu, Yuhuai and Rabe, Markus N. and Hutchins, DeLesley and Szegedy, Christian},
  booktitle={ICLR},
  year={2022},
  note={arXiv:2203.08913}
}

@inproceedings{khandelwal2019knnlm,
  title={Generalization through Memorization: Nearest Neighbor Language Models},
  author={Khandelwal, Urvashi and Levy, Omer and Jurafsky, Dan and Zettlemoyer, Luke and Lewis, Mike},
  booktitle={ICLR},
  year={2020},
  note={arXiv:1911.00172}
}

@inproceedings{hendrycks2021mmlu,
  title={Measuring Massive Multitask Language Understanding},
  author={Hendrycks, Dan and Burns, Collin and Basart, Steven and Zou, Andy and Mazeika, Mantas and Song, Dawn and Steinhardt, Jacob},
  booktitle={ICLR},
  year={2021},
  note={arXiv:2009.03300}
}

@article{srivastava2022bigbench,
  title={Beyond the Imitation Game: Quantifying and Extrapolating the Capabilities of Language Models},
  author={Srivastava, Aarohi and Rastogi, Abhinav and others},
  journal={arXiv preprint arXiv:2206.04615},
  year={2022}
}

@article{liang2022helm,
  title={Holistic Evaluation of Language Models},
  author={Liang, Percy and Bommasani, Rishi and others},
  journal={arXiv preprint arXiv:2211.09110},
  year={2022}
}

@inproceedings{lin2022truthfulqa,
  title={TruthfulQA: Measuring How Models Mimic Human Falsehoods},
  author={Lin, Stephanie and Hilton, Jacob and Evans, Owain},
  booktitle={ACL (Long Papers)},
  year={2022},
  note={arXiv:2109.07958}
}

@article{shoeybi2019megatron,
  title={Megatron-LM: Training Multi-Billion Parameter Language Models Using Model Parallelism},
  author={Shoeybi, Mohammad and Patwary, Mostofa and Puri, Raul and LeGresley, Patrick and Casper, Jared and Catanzaro, Bryan},
  journal={arXiv preprint arXiv:1909.08053},
  year={2019}
}

@inproceedings{rajbhandari2020zero,
  title={{ZeRO}: Memory Optimizations Toward Training Trillion Parameter Models},
  author={Rajbhandari, Samyam and Rasley, Jeff and Ruwase, Olatunji and He, Yuxiong},
  booktitle={SC ’20: International Conference for High Performance Computing, Networking, Storage and Analysis},
  year={2020}
}

@inproceedings{hinton2015distill,
  title={Distilling the Knowledge in a Neural Network},
  author={Hinton, Geoffrey and Vinyals, Oriol and Dean, Jeff},
  booktitle={NIPS Deep Learning Workshop},
  year={2015},
  note={arXiv:1503.02531}
}

@article{dettmers2022llmint8,
  title={{LLM.int8()}: 8-bit Matrix Multiplication for Transformers at Scale},
  author={Dettmers, Tim and Lewis, Mike and Belkada, Younes and Zettlemoyer, Luke},
  journal={arXiv preprint arXiv:2208.07339},
  year={2022},
  note={Also in Proc. ACM (DOI:10.5555/3600270.3602468)}
}

@article{frantar2022gptq,
  title={{GPTQ}: Accurate Post-Training Quantization for Generative Pre-trained Transformers},
  author={Frantar, Elias and Ashkboos, Saleh and Hoefler, Torsten and Alistarh, Dan},
  journal={arXiv preprint arXiv:2210.17323},
  year={2022}
}

@article{dettmers2023qlora,
  title={{QLoRA}: Efficient Finetuning of Quantized LLMs},
  author={Dettmers, Tim and Pagnoni, Artidoro and Holtzman, Ari and Zettlemoyer, Luke},
  journal={arXiv preprint arXiv:2305.14314},
  year={2023}
}

@article{zhou2023webarena,
  title={WebArena: A Realistic Web Environment for Building Autonomous Agents},
  author={Zhou, Shuyan and Xu, Frank F. and Zhu, Hao and Zhou, Xuhui and Lo, Robert and others},
  journal={arXiv preprint arXiv:2307.13854},
  year={2023}
}

@article{liu2023agentbench,
  title={AgentBench: Evaluating LLMs as Agents},
  author={Liu, Xiao and Yu, Hao and Zhang, Hanchen and Xu, Yifan and Lei, Xuanyu and others},
  journal={arXiv preprint arXiv:2308.03688},
  year={2023}
}

@article{wang2023voyager,
  title={Voyager: An Open-Ended Embodied Agent with Large Language Models},
  author={Wang, Guanzhi and Xie, Yuqi and Jiang, Yunfan and Mandlekar, Ajay and Xiao, Chaowei and Zhu, Yuke and Fan, Linxi and Anandkumar, Anima},
  journal={arXiv preprint arXiv:2305.16291},
  year={2023}
}

@misc{AIJobs-Applied-Ads-2023,
  author       = {{Amazon.com}},
  title        = {Applied Scientist, Advertising (Job Listing)},
  howpublished = {AI Jobs posting (London, UK)},
  year         = {2023},
  note         = {Preferred qualifications include deep learning frameworks and distributed systems},
  url          = {https://aijobs.net/job/1435356-applied-scientist-advertising/},
  urldate      = {2025-07-17}
}

@misc{MetaCareers-GenAI-2025,
  author       = {{Meta}},
  title        = {Research Engineer, Generative AI (Job Listing)},
  howpublished = {Meta Careers (LinkedIn Job Post)},
  year         = {2025},
  note         = {Highlights need for generative AI and NLP experience, model evaluation, etc.},
  url          = {https://www.linkedin.com/jobs/view/3584278795/},
  urldate      = {2025-07-17}
}

\end{document}